\documentclass[11pt,a4paper]{article}

\usepackage[nohead, nomarginpar, margin=1in, foot=.25in]{geometry}

\usepackage{graphicx}

\usepackage{csquotes}

\usepackage{amsmath}

\usepackage[T1]{fontenc}
\usepackage[utf8]{inputenc}
\usepackage{authblk}

\usepackage[LGRgreek]{mathastext}

\usepackage{rotating}
\usepackage{graphicx}
\usepackage{amssymb}
\usepackage{epstopdf}
\usepackage{subfigure}
\usepackage{caption}
\usepackage{lipsum}
\usepackage[pagewise]{lineno}

\date{}

\newcommand{\sn}{\sqrt{s_{\rm{NN}}}}
\newcommand{\s}{\sqrt{s}}
\newcommand{\pt}{\textit{p}_{{T}}}

\newcommand{\dnchdeta}{\langle d \textit{N}_{ch} / d {\eta} \rangle} 

\begin{document}
\makeatletter
  \def\title@font{\Large\bfseries}
  \let\ltx@maketitle\@maketitle
  \def\@maketitle{\bgroup%
    \let\ltx@title\@title%
    \ltx@maketitle%
  \egroup}
\makeatother

\title{\bf \hspace{0.7cm} Hadron production in pp and p-Pb collisions: \newline A mass dependent phenomenon }
\author[1,2,\footnote{ssarita.2006@gmail.com}]{S. Sahoo}
\author[$$]{R. C. Baral}
\author[1,2]{P. K. Sahu}
\author[2]{M. K. Parida}

\affil[1] {Institute of Physics, HBNI, Bhubaneswar, India}
\affil[2] {CETMS, Siksha `O' Anusandhan Deemed to be University, Bhubaneswar, India}

\maketitle

\begin{abstract}
The mass dependence plays a significant role in the yield enhancement or suppression of hadrons in pp and p--Pb collisions at the LHC energies.
This has been observed by parameterizing the variation of yield ratios between any two hadrons with event charged-particle multiplicity using a single empirical function. 
We notice that this variation is independent of all quantum numbers and solely depends on masses of hadrons and masses of their valence quarks.
The function shows that the amount of quark deconfinement increases with event multiplicity, and the quark coalescence favours more the production of heavier hadrons compared to lighter ones.
\end{abstract}

\section{Introduction}
Heavy-ion collisions at ultra relativistic energies produce a new form of QCD matter characterized by the deconfined state of quarks and gluons, known as the QGP \cite{s_QGP1,s_QGP2}. Measurements of the production of identified particles in high-energy nucleus-nucleus (AA) collisions, relative to proton-proton (pp) or proton-nucleus (pA) collisions, provide information about the dynamics of this hot and dense matter. In pp and pA collisions, the relative contributions to hadron production through different hadronization mechanisms change with event charged-particle multiplicity. 
Quark (re-)combination mechanism describes the formation of hadrons at hadronization as follows \cite{qCols1, qCols2, qCols3, qCols4, qCols5}: A quark and an antiquark neighboring in phase space form a meson and three quarks (antiquarks) form a baryon (antibaryon).
High transverse momentum ($\pt$) quarks can be produced in hard (perturbative) partonic scattering processes like flavor creation, flavor excitation and gluon splitting. These quarks tend to dominate the production of high $\pt$ hadrons through fragmentation \cite{frag1,frag2,frag3}. At low $\pt$, the non-perturbative processes like pair production and string fragmentation dominate the hadron production.

The non-valence strange quarks ($\it{s}$) are sufficiently light ($\sim$96 MeV/$\it{c}$$^2$) \cite{PDG} to be abundantly created during the course of collisions. In spite of that, these are many times heavier than non-valence $\it{u}$ and $\it{d}$ quarks. 
The production of hadrons containing $\it{s}$ quark(s) appears to be significantly suppressed in smaller systems \cite{StrngSup1, StrngSup2, StrngSup3}.
According to the Statistical Hadronization Model \cite{SHM1, SHM2,SHM3}, the suppression of relative strangeness production in elementary collisions with respect to heavy-ion collisions is mainly an effect of the decrease of the global volume from heavy-ion to elementary collisions. This phenomenon is known as strangeness canonical suppression that requires conservation of strangeness. However, canonical suppression is not enough to account for strangeness enhancement from pp to heavy-ion collisions. This is demonstrated by neutral mesons like $\Phi$ meson, which does not suffer canonical suppression but are relatively more abundant in heavy-ion collisions \cite{PhiPbPb1, PhiPbPb2}. 
In high-multiplicity pp collisions, the integrated yields of strange and multi-strange particles, relative to $\pi$, increase significantly with event charged-particle multiplicity density ($\dnchdeta$) \cite{pp7TeV, pp7TeV2019}. The enhancement is more pronounced for multi-strange baryons. 
\newline
\newline
In this paper, we present the enhancements or suppressions in the yield ratios between any two hadrons (mesons or baryons, strange or non-strange particles) with event charged-particle multiplicity by using ALICE published data in pp collisions at $\s$ = 7 TeV \cite{pp7TeV, pp7TeV2019} and in p--Pb collisions at $\sn$ = 5.02 TeV \cite{pPb1,pPb2}. We describe the relative yield ratios by fitting the published data with a parameterized function. We find that the available data follow this hypothesis: the variations in yield ratios depend on masses of hadrons and masses of their valence quarks, and this is independent of any quantum number. The function suggests the dominance of different hadron production mechanisms varies with $\dnchdeta$.
%
%

\section{Results and Discussions} \label{Results and Discussions}
\begin{figure}[!b]
\centering
\includegraphics[scale=0.41]{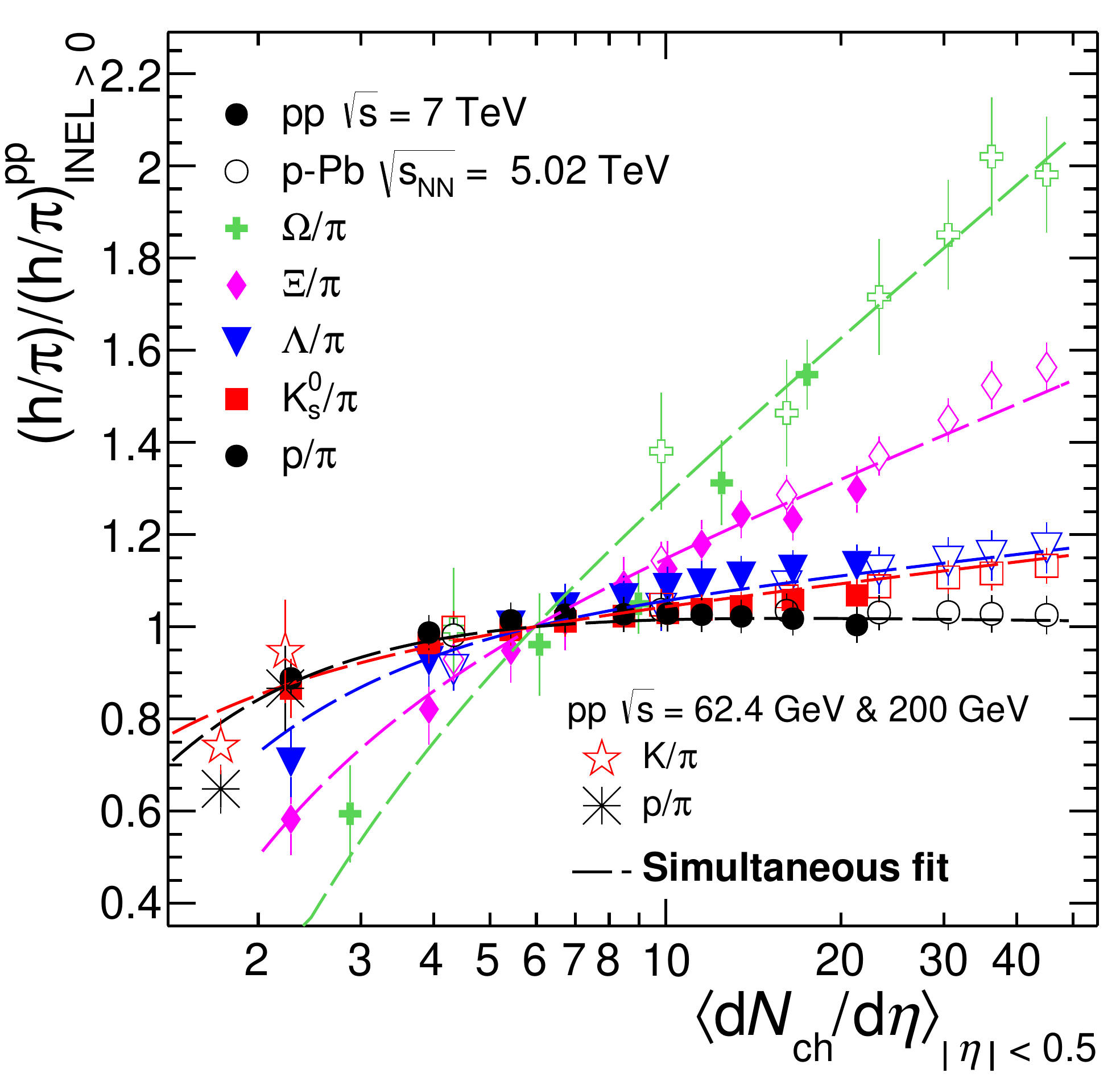}
\caption{\small Yield ratios of non-strange (p), single strange (K$_s^0$, $\Lambda$) and multi-strange ($\Xi$, $\Omega$) hadrons to $\pi$, scaled by the respective values measured in the inclusive (INEL > 0) pp collisions, and plotted against $\dnchdeta$. The results are shown for pp collisions at $\s$ = 7 TeV \cite{pp7TeV,pp7TeV2019} and  p--Pb collisions at $\sn$ = 5.02 TeV  \cite{pPb1, pPb2}. The PHENIX published results for K/$\pi$ and p/$\pi$ in pp collisions at $\s$ = 62.4 and 200 GeV are shown by open stars and asterisks symbols, respectively \cite{phenix} . 
Statistical and uncorrelated systematic uncertainties are added in quadrature and are shown by vertical bars. The lines represent a simultaneous fit of the results with the empirical function in equation (1).}
\label{fig:simultaneousfit}
\end{figure}
%

The strange and multi-strange baryon yields in Pb--Pb collisions have been shown to exhibit an enhancement relative to pp reactions \cite{PbPb}. The same can be seen for high multiplicity events relative to low multiplicity events in pp collisions at LHC energies \cite{pp7TeV,pp7TeV2019}. Figure~\ref{fig:simultaneousfit} shows the multiplicity dependence of primary yield ratios of hadrons to $\pi$, scaled by the respective values measured in the inclusive (INEL > 0) pp collisions. Here a primary particle is defined as a particle created in the collision, but not coming from a weak decay. The measurements were performed for events having at least one charged particle produced in the pseudorapidity interval $|\eta|$ < 1.0 (i.e. INEL > 0). 
For brevity, $\pi^+ + \pi^-$, p + $\overline{p}$, $\Lambda + \overline{\Lambda}$, $\Xi^- + \overline{\Xi}^{_+}$ and $\Omega^- + \overline{\Omega}^{_+}$ are denoted as $\pi$, p, $\Lambda$, $\Xi$ and $\Omega$, respectively.
The hadrons in the numerator are non-strange (p), single strange (K$_s^0$, $\Lambda$) and multi-strange ($\Xi$, $\Omega$) particles. The published experimental data points are shown by markers (see legends in the Fig. 1), and these are in pp collisions at $\s$ = 7 TeV \cite{pp7TeV,pp7TeV2019} and in p--Pb collisions at $\sn$ = 5.02 TeV \cite{pPb1, pPb2}. The measurements were performed at mid-rapidity $|{\it y}|$ < 0.5. The details of information on data can be found in Refs.~\cite{pp7TeV, pPb1, pPb2}  and references therein. The figure shows that the normalized ratios increase with  $\dnchdeta$. The ratio between two non-strange hadrons (p/$\pi$) is close to unity over all $\dnchdeta$ except at the lowest point. 
One can observe that $\Lambda/\pi$ values continuously stay above (below) K$_s^0/\pi$ (though both the ratios are consistent within uncertainty) for $\dnchdeta$ above (below)  $\dnchdeta^{pp}_{INEL > 0}$, which is the charged-particle multiplicity density in (INEL > 0) pp collisions. Also, the hadron ratios at the lowest $\dnchdeta$ appear to be consistently going out of the trend (see the parameterized function fit in Ref. \cite{pp7TeV}). 
The data indicate that the larger is the mass difference between numerator and denominator, the higher is the suppression in the low $\dnchdeta$ region. 
To make this indication more promising, we have shown PHENIX published results for K/$\pi$ and p/$\pi$ in pp collisions at $\s$ = 62.4 and 200 GeV in Fig.~\ref{fig:simultaneousfit} at their respective $\dnchdeta$ \cite{phenix}.
These observations suggest that there may be a mass effect in the enhancement and/or suppression, and the dominant mechanisms for the hadron production may differ with $\dnchdeta$.
We have parameterized a function to describe this behaviour and examined how effectively this function can fit these 5 sets of data points simultaneously. Every line in Fig. \ref{fig:simultaneousfit} represents the curve predicted by our function that fits the corresponding data points.
The empirical function is in the form
%
\begin{figure}[!b]
\centering
\includegraphics[scale=0.41]{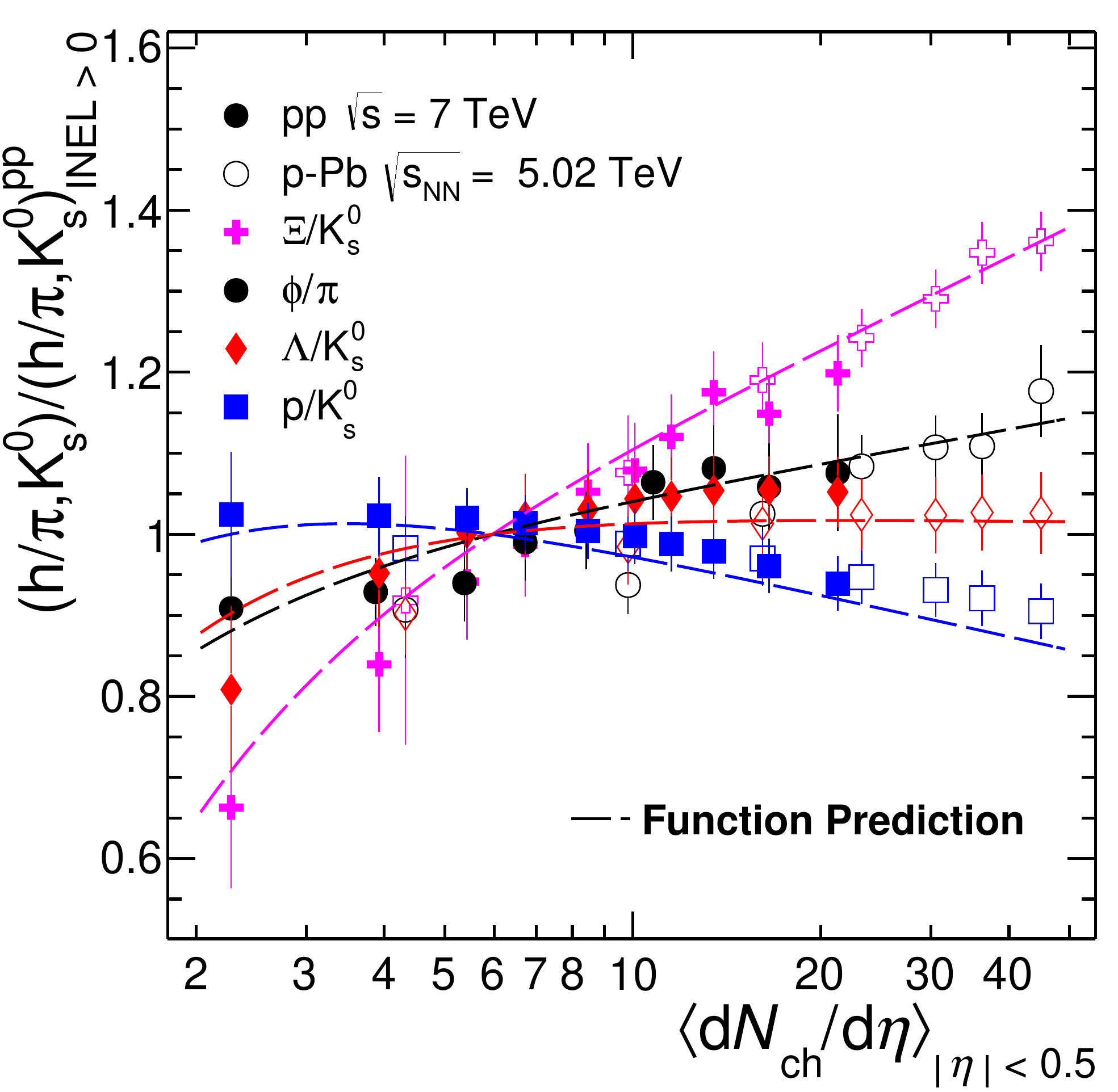}
\caption{\small Particle yield ratios scaled by the respective values measured in the inclusive (INEL > 0) pp collisions, and plotted against $\dnchdeta$. The results are shown for pp collisions at $\s$ = 7 TeV \cite{pp7TeV,pp7TeV2019} and p--Pb collisions at $\sn$ = 5.02 TeV \cite{pPb1, pPb2}. Statistical and uncorrelated systematic uncertainties are added in quadrature and are shown by vertical bars. The lines represent predictions of the results by the empirical function in equation (1).}
\label{fig:Predictions}
\end{figure}
%
\begin{equation}
\begin{split}
\frac{h/\pi}{(h/\pi)^{pp}_{INEL > 0}} = 1 & + a \left((M_{q,N}){^b}-(M_{q,D})^b\right) log\left[ \frac{\langle d \textit{N}_{ch} / d {\eta} \rangle}  {\langle d \textit{N}_{ch} / d {\eta} \rangle ^{pp}_{INEL > 0}} \right] \\
& + c \left(M_{h,N} - M_{h,D} \right) log\left[ \frac{\langle d \textit{N}_{ch} / d {\eta} \rangle}  {\langle d \textit{N}_{ch} / d {\eta} \rangle ^{pp}_{INEL > 0}} \right] \bigg/ \langle d \textit{N}_{ch} / d {\eta} \rangle , 
\end{split}
\end{equation}
where   $M_{q,N}$ ($M_{q,D}$) is twice the sum of masses of valence quarks of the hadron in the numerator 
(denominator), and $M_{h,N}$ ($M_{h,D}$) is twice the mass of hadron in the numerator (denominator) \cite{PDG}. All the mass values are taken in GeV/$\it{c}$$^2$ and the values used  for $\it u$, $\it d$ and $\it s$ quarks are 0.0022 GeV/$\it{c}$$^2$, 0.0047 GeV/$\it{c}$$^2$ and 0.096 GeV/$\it{c}$$^2$, respectively \cite{PDG}. 
$(h/\pi)^{pp}_{INEL > 0}$ is the measured hadron to $\pi$ ratio in (INEL > 0) pp collisions, and a, b, c are free parameters having appropriate units. The fit describes available data well with $\chi^2$/ndf = 0.38 for the best fit parameter values a = 1.35 $\pm$ 0.13, b = 1.84 $\pm$ 0.11 and c = 0.18 $\pm$ 0.034. 
There are two parts in the empirical function, which represents two different behaviours of yield enhancement. The first part depends on the masses of valence quarks, and it contributes substantially towards higher $\dnchdeta$. The second part depends on the hadron mass, and it contributes largely towards lower $\dnchdeta$. This may indicate that hadron production towards higher $\dnchdeta$  is largely from quark coalescence and recombination mechanism \cite{qCols1, qCols2, qCols3, qCols4, qCols5}, while in the lower $\dnchdeta$ region, most hadrons come from fragmentation like processes where hadrons are mainly produced by excited string fragmentations or high $\pt$ partonic jets \cite{frag1,frag2, frag3}. In general, the coalescence models have been qualitatively successful in describing the increase of the strange baryons to meson ratios \cite{saritaproced}. At the LHC energies, data show net baryon density vanishes at mid-rapidity \cite{netbar1,netbar2,netbar3,netbar4}. So quarks form in pairs ($ \it{q\overline{q}}$) of same flavour for the conservation of baryon number. It requires 2 quark pairs ($ \it{u\overline{u}}$ \& $ \it{d\overline{d}}$ ) to form a single $\pi^+$ ($ \it{u\overline{d}}$). Same process is followed for all hadrons except for few neutral mesons, especially $\Phi$ mesons. Similar way, a string has to fragment into hadron and  its anti-hadron. This makes us double the masses in the formula. 
Therefore, the function is expected to deviate from the data with the increase of net event baryon density. This is shown in Fig.~\ref{fig:simultaneousfit} by K/$\pi$ and p/$\pi$ ratios in pp collisions at $\s$ = 62.4 GeV.

Figure \ref{fig:Predictions} shows the published data (by markers) for different hadrons ($\Xi$, $\Phi$, $\Lambda$, p) to $\pi$ and K$_s^0$ ratios, scaled by the respective values measured in the inclusive (INEL > 0) pp collisions, as a function of $\dnchdeta$ in pp collisions \cite{pp7TeV,pp7TeV2019} and p--Pb collisions \cite{pPb1, pPb2}.
The lines represent the function predictions for the respective ratios (follow the legend of Fig. \ref{fig:Predictions}). 
The function describes data to a great extent both qualitatively and quantitatively. The data show $\Phi$ meson behaves like a single strange particle. We do not need to double the masses for $\Phi$ meson in the function to describe its behaviour. Therefore, the function says that the neutral mesons are likely to be less affected by strangeness (canonical) suppression, and this can be seen towards lower $\dnchdeta$ region.
Despite the same strange content, $\Lambda$/K$_s^0$ ratio, like p/$\pi$ in Fig. \ref{fig:simultaneousfit}, shows suppression towards lower $\dnchdeta$. The p/K$_s^0$ ratio agrees with unity towards lower $\dnchdeta$. This may indicate a cancelling effect between higher p mass and heavier $\it s$ quark in K$_s^0$. The normalized $\Xi$/K$_s^0$ ratio stays below normalized $\Xi$/$\pi$ ratio and above normalized $\Lambda$/$\pi$ ratio. 
All these behaviours of experimental data are well described by the function. To know how well the functions describe the individual set of ratios, we have calculated $\chi^2$/ndf for each function. Each $\chi^2$/ndf is below 1 except for $\Omega/\pi$, which is 1.07. 
These values are better as compared to those $\chi^2$/ndf values, which are obtained if one uses the empirical function proposed by ALICE \cite{pp7TeV}. 
\begin{figure}[!]
\centering
\includegraphics[scale=0.71]{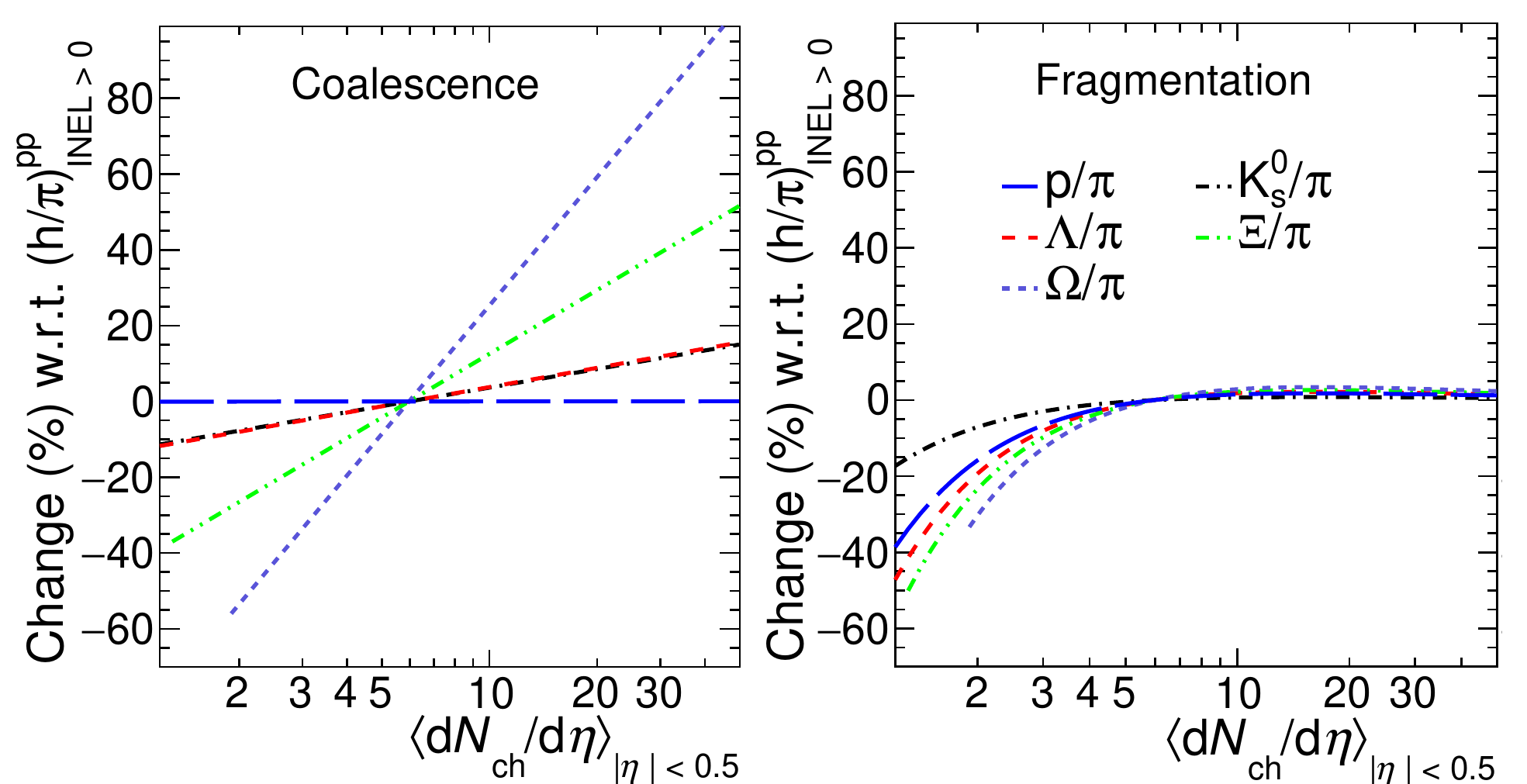}
\caption{\small Plot in the left (right) shows the change ($\%$) in the ${\it h}/\pi$ ratio from $ \left( {\it h}/\pi \right )^{pp}_{INEL > 0}$ as a function of $\dnchdeta$, and that comes from the first (second) part of the function.}
\label{fig:M_SFracToNet}
\end{figure}

The two parts in the empirical function give suppression or enhancement in the yield ratios of hadrons independently, and it is shown in Figure \ref{fig:M_SFracToNet}. This shows the changes (\%) in the ${\it h}/\pi$ ratios from $\left< {\it h}/\pi \right >^{pp}_{INEL > 0}$  as a function of $\dnchdeta$. The plot on the left (right) shows that the change comes from the first (second) part of function.
This shows coalescence of quarks favours the production of heavier hadrons relative to lighter ones with increasing $\dnchdeta$, and fragmentation suppresses significantly production of heavier hadrons relative to lighter ones towards lower $\dnchdeta$.
The first part of the function shows similar enhancement and suppression in K$_s^0/\pi$ and $\Lambda/\pi$ ratios, but the second part shows their suppressions towards low $\dnchdeta$ are hadron mass dependent.
Plots in the left-panel and the
right-panel of Fig. \ref{fig:M_SFracToNet} show there is a hierarchy in suppression or enhancement that depends upon the masses of valence quarks and masses of hadrons, respectively.
The function may not calculate the fraction of yields produced in either of the hadron production mechanisms and also, may not be applied for event with non-zero net baryon density. We may expect from the function to describe the suppression or enhancement, if any, in charm hadron production relative to lighter hadrons as well.



\section{Conclusion} 
\label{Conclusion}
\noindent 
To summaries, we have parameterized a function to describe multiplicity dependent relative yield enhancement of hadrons produced in pp collisions at $\s$ = 7 TeV and p--Pb collisions at $\sn$ = 5.02 TeV. The function depends only on masses of hadrons and masses of valence quarks, and it shows that the enhancement does not depend upon any quantum number. 
A possible mass hierarchy has been seen in hadron production and their relative yield enhancement. 
This may point towards a common underlying physics mechanism for hadron production in these collision systems. 
The function depicts graphically the contributions of hadronization mechanisms to hadron yields, where it shows quark coalescence mechanism dominates over fragmentation for high multiplicity events, and the former mechanism favours the production of hadrons containing heavier quark(s). 
This observation is very interesting in view of that the partonic degree of freedom gradually playing a crucial role in particle production, especially for the multi-strange particles in high multiplicity pp collisions. This may be conjectured to be a signal of formation of a partonic phase in pp collisions at LHC energies. 
In high multiplicity events, it further shows that fragmentation  mechanism equally favours production of strange and non-strange hadrons, and it is able to describe the sign of suppression in yields of heavier hadron relative to lighter ones for low multiplicity events. 
The function agrees to accept that $\Phi$ meson is likely to be less affected by strangeness canonical suppression.
Further studies extending to heavy flavour production in pp collisions are essential. It would confirm predictive
capability of the function in case of charm production.


\end{document}